\documentclass[aps,prd,preprint,superscriptaddress,preprintnumbers,nofootinbib,tightenlines,floatfix]{revtex4-1}
\usepackage{graphics}
\usepackage{color}
\usepackage{graphicx}
\usepackage{amssymb}
\usepackage{amsmath}
\usepackage{epstopdf}
\usepackage{xspace}

\usepackage{mathbbol}
\usepackage{wasysym} 
\usepackage{bbm}     
 
\usepackage{rotating}
\usepackage{dcolumn}
\usepackage{bm}
\usepackage{longtable}
\usepackage{multirow}
\usepackage{enumitem}

\usepackage{ifpdf}
\ifpdf
\usepackage[pdftex]{hyperref}
\else
\usepackage[hypertex]{hyperref}
\fi

\hypersetup{
 pdftitle={},%
 pdfauthor={},%
 pdfsubject={},%
 pdfkeywords={},%
 pdfstartview={},%
 bookmarksopen=true, breaklinks=true, debug=true,
 colorlinks=true, linkcolor=blue, citecolor=red, urlcolor=red,
 hyperfigures=true
}

\def \brf{{\cal B}}
\def \bea{\begin{eqnarray}}
\def \beq{\begin{equation}}

\def \eea{\end{eqnarray}}
\def \eeq{\end{equation}}
\def \bea{\begin{eqnarray}}
\def \beq{\begin{equation}}

\def \eea{\end{eqnarray}}
\def \eeq{\end{equation}}

\usepackage{relsize}
\def\babar{\mbox{\slshape B\kern-0.1em{\smaller A}\kern-0.1em
    B\kern-0.1em{\smaller A\kern-0.2em R}}}
\def\mev{~MeV}

\def\mevcc{\mev/$c^2$}
\def\gev{~GeV}
\def\gevc{\gev/$c$}
\def\gevcc{\gev/$c^2$}
\def\bbbar{b\bar{b}}
\def\ccbar{c\bar{c}}
\def\bbccbar{\bbbar~[\ccbar]}
\def\hb{h_b}
\def\hc{h_c}
\def\hbc{h_{b[c]}}

\def\etab{\eta_b}
\def\etac{\eta_c}
\def\etabc{\eta_{b[c]}}

\def\chibj{\chi_{bJ}}
\def\chicj{\chi_{cJ}}
\def\chibcj{\chi_{b[c]J}}

\def\chict{\chi_{c2}}

\def\ut{\Upsilon(3S)}
\def\pp{\psi(2S)}
\def\utpp{\ut~[\pp]}
\def\piz{{\pi^0}}
\def\tg{{\gamma\gamma}}
\def\pizgg{\piz \to \tg}
\def\etagg{\eta \to \tg}
\def\utphb{\ut \to \piz \hb}
\def\ppphc{\pp \to \piz \hc}
\def\butphb{\brf(\utphb)}
\def\bppphc{\brf(\ppphc)}
\def\utppphbc{\ut [\pp]\to\piz\hbc}
\def\butppphbc{\brf(\ut [\pp]\to\piz\hbc)}

\def\bhbgeb{\brf(\hb\to\gamma\etab)}
\def\bhcgec{\brf(\hc\to\gamma\etac)}
\def\bhbcgebc{\brf(\hbc\to\gamma\etabc)}

\def\gchibj{\gamma \chibj}

\def\gchict{\gamma \chict}
\def\gchicj{\gamma \chicj}
\def\utgchibj{\ut \to \gchibj}

\def\ppgchict{\pp \to \gchict}
\def\ppgchicj{\pp \to \gchicj}

\def\butgchibj{\brf(\utgchibj)}
\def\bppgchicj{\brf(\ppgchicj)}

\def\gchibcj{\gamma \chibcj}
\def\utppgchibcj{\ut [\pp] \to \gchibcj}

\def\recmas{M_{\rm rec}(\piz)}

\def\aca{|\cos\alpha|}
\def\acamax{\aca_{\rm max}}

\def\Nres{N_{\rm res}}
\def\Nevt{N_{\rm evt}}

\begin{document}

\title{\boldmath Branching fractions for $\Upsilon(3S)\to\pi^0h_b$
and $\psi(2S)\to\pi^0h_c$}

\preprint{CLNS 11/2079} 
\preprint{CLEO 11-05}  

\author{J.~Y.~Ge}
\author{D.~H.~Miller}
\author{I.~P.~J.~Shipsey}
\author{B.~Xin}
\affiliation{Purdue University, West Lafayette, Indiana 47907, USA}
\author{G.~S.~Adams}
\author{J.~Napolitano}
\affiliation{Rensselaer Polytechnic Institute, Troy, New York 12180, USA}
\author{K.~M.~Ecklund}
\affiliation{Rice University, Houston, Texas 77005, USA}
\author{J.~Insler}
\author{H.~Muramatsu}
\author{C.~S.~Park}
\author{L.~J.~Pearson}
\author{E.~H.~Thorndike}
\affiliation{University of Rochester, Rochester, New York 14627, USA}
\author{S.~Ricciardi}
\affiliation{STFC Rutherford Appleton Laboratory, Chilton, Didcot, Oxfordshire, OX11 0QX, UK}
\author{C.~Thomas}
\affiliation{University of Oxford, Oxford OX1 3RH, UK}
\affiliation{STFC Rutherford Appleton Laboratory, Chilton, Didcot, Oxfordshire, OX11 0QX, UK}
\author{M.~Artuso}
\author{S.~Blusk}
\author{R.~Mountain}
\author{T.~Skwarnicki}
\author{S.~Stone}
\author{L.~M.~Zhang}
\affiliation{Syracuse University, Syracuse, New York 13244, USA}
\author{G.~Bonvicini}
\author{D.~Cinabro}
\author{A.~Lincoln}
\author{M.~J.~Smith}
\author{P.~Zhou}
\author{J.~Zhu}
\affiliation{Wayne State University, Detroit, Michigan 48202, USA}
\author{P.~Naik}
\author{J.~Rademacker}
\affiliation{University of Bristol, Bristol BS8 1TL, UK}
\author{D.~M.~Asner}
\altaffiliation[Now at: ]{Pacific Northwest National Laboratory, Richland, WA 99352}
\author{K.~W.~Edwards}
\author{K.~Randrianarivony}
\author{G.~Tatishvili}
\altaffiliation[Now at: ]{Pacific Northwest National Laboratory, Richland, WA 99352}
\affiliation{Carleton University, Ottawa, Ontario, Canada K1S 5B6}
\author{R.~A.~Briere}
\author{H.~Vogel}
\affiliation{Carnegie Mellon University, Pittsburgh, Pennsylvania 15213, USA}
\author{P.~U.~E.~Onyisi}
\author{J.~L.~Rosner}
\affiliation{University of Chicago, Chicago, Illinois 60637, USA}
\author{J.~P.~Alexander}
\author{D.~G.~Cassel}
\author{S.~Das}
\author{R.~Ehrlich}
\author{L.~Gibbons}
\author{S.~W.~Gray}
\author{D.~L.~Hartill}
\author{B.~K.~Heltsley}
\author{D.~L.~Kreinick}
\author{V.~E.~Kuznetsov}
\author{J.~R.~Patterson}
\author{D.~Peterson}
\author{D.~Riley}
\author{A.~Ryd}
\author{A.~J.~Sadoff}
\author{X.~Shi}
\author{W.~M.~Sun}
\affiliation{Cornell University, Ithaca, New York 14853, USA}
\author{J.~Yelton}
\affiliation{University of Florida, Gainesville, Florida 32611, USA}
\author{P.~Rubin}
\affiliation{George Mason University, Fairfax, Virginia 22030, USA}
\author{N.~Lowrey}
\author{S.~Mehrabyan}
\author{M.~Selen}
\author{J.~Wiss}
\affiliation{University of Illinois, Urbana-Champaign, Illinois 61801, USA}
\author{J.~Libby}
\affiliation{Indian Institute of Technology Madras, Chennai, Tamil Nadu 600036, India}
\author{M.~Kornicer}
\author{R.~E.~Mitchell}
\author{C.~M.~Tarbert}
\affiliation{Indiana University, Bloomington, Indiana 47405, USA }
\author{D.~Besson}
\affiliation{University of Kansas, Lawrence, Kansas 66045, USA}
\author{T.~K.~Pedlar}
\affiliation{Luther College, Decorah, Iowa 52101, USA}
\author{D.~Cronin-Hennessy}
\author{J.~Hietala}
\affiliation{University of Minnesota, Minneapolis, Minnesota 55455, USA}
\author{S.~Dobbs}
\author{Z.~Metreveli}
\author{K.~K.~Seth}
\author{A.~Tomaradze}
\author{T.~Xiao}
\affiliation{Northwestern University, Evanston, Illinois 60208, USA}
\author{L.~Martin}
\author{A.~Powell}
\author{G.~Wilkinson}
\affiliation{University of Oxford, Oxford OX1 3RH, UK}
\collaboration{CLEO Collaboration}
\noaffiliation

\date{June 17, 2011}

\begin{abstract} 
Using $e^+e^-$ collision data corresponding to $5.88\times 10^6~\Upsilon(3S)$ 
[$25.9\times 10^6~\psi(2S)$] decays and acquired by the CLEO~III [CLEO-c] 
detectors operating at 
the Cornell Electron Storage Ring,
we study the single-pion transitions from $\Upsilon(3S)~[\psi(2S)]$ 
to the respective
spin-singlet states $h_{b[c]}$. Utilizing only the momentum of suitably
selected transition-$\pi^0$ candidates, we obtain the upper limit 
${\cal B}(\Upsilon(3S)\to\pi^0h_b) < 1.2\times 10^{-3}$ 
at 90\% confidence level,
and measure $\bppphc =(9.0\pm1.5\pm1.3)\times 10^{-4}$. 
Signal sensitivities are enhanced by excluding very asymmetric  $\pi^0\to\gamma\gamma$  candidates.
\end{abstract}

\maketitle

Hadronic transitions between heavy 
quarkonium
states provide
a rich tableau of opportunities for experimental and theoretical
investigations~\cite{hqppo}. Such transitions can supply important production
mechanisms for the lower-lying state, thereby allowing measurement
of its mass, width, and decay channels, and can also 
enable nonperturbative QCD calculations
of transition rates to confront experiment. In particular,
observation and exploration of the $\bbccbar$
spin-singlet states $\hbc$ have depended strongly upon
hadronic transitions from vector quarkonia produced in $e^+e^-$ collisions. 
The $\hbc$ states\footnote{Throughout this article, lowest radial
excitations are implied unless explicitly indicated.} were both
expected and measured to have masses near the
spin-weighted averages of the respective spin-triplet
states $\chibcj$. Observations of $\ppphc$ were reported first
by CLEO~\cite{CLEOhc} and later by BESIII~\cite{BEShc}. Evidence
for $\utphb$ has been reported by \babar~\cite{BaBarhb}. Charged dipion
transitions from higher-lying vector states were first shown to
produce $\hc$ by CLEO~\cite{pipihc} and, later, 
both $\hb(1P)$ and $\hb(2P)$ by Belle~\cite{Bellehb}.

In this article we report on attempts to measure
the branching fractions for the isospin-violating 
single-$\piz$ transitions from $\utpp$ to $\hbc$. Previously,
\babar~\cite{BaBarhb} measured the product branching
fraction $\butphb\times\bhbgeb=(4.3\pm1.1\pm0.9)\times10^{-4}$,
finding evidence for a signal with significance of $3.3\sigma$, 
consistent with predictions 
$\butphb\approx (1.0{\mathrm -}1.6) \times 10^{-3}$~\cite{hbprod,hbprivate}
and $\bhbgeb=41$\%~\cite{hbmodel}.  Meanwhile, BESIII~\cite{BEShc}
measured $\bppphc=(8.4\pm1.3\pm1.0)\times 10^{-4}$ (in conjunction with
a simultaneously determined $\bhcgec=(54.3\pm6.7\pm5.2)\%$), compatible with
both potential model estimates~\cite{Kuang:2002hz} and
a non-relativistic effective field theory
prediction~\cite{Guo:2010zk} (see also \cite{Burns:2011fu}) which found charmed
meson loop contributions to mass shifts to be atypically small.

The two data sets used in this work were collected in $e^+e^-$
collisions at the Cornell Electron Storage Ring,
at the
center-of-mass energies of the $\Upsilon(3S)$ and $\psi(2S)$ resonances,
corresponding to $(5.88 \pm 0.12)\times 10^6$ $\Upsilon(3S)$ 
decays~\cite{Nups} and $(25.9 \pm 0.5)\times 10^6$ $\psi(2S)$ 
decays~\cite{npsi2s}. Events were recorded using the CLEO~III and 
CLEO-c detectors for $\Upsilon(3S)$ and $\psi(2S)$ datasets, respectively.
Both configurations are equipped with an electromagnetic calorimeter
consisting of 7784 thallium-doped cesium iodide crystals and covering
93\% of solid angle, initially installed in the CLEO~II~\cite{CLEO2}
detector configuration. The energy resolution of the crystal
calorimeter is 5\% (2.2\%) for 0.1\gev\ (1\gev) photons.
Calorimeter angular resolution is $\sim$10 (5) mrad at $E_\gamma =$
100 MeV (1 GeV), and does not significantly contribute to 
$\gamma\gamma$ mass resolution for the soft $\pi^0$ candidates
considered in this analysis.
The CLEO~III
tracking system~\cite{CLEO3trk} consists of
a silicon strip vertex detector and a large drift chamber; 
a six-layer wire vertex detector replaced the silicon in the
CLEO-c configuration~\cite{CLEOc}.  The trackers achieve charged particle
momentum resolutions of 0.35\% and 0.6\% at 1\gevc\ in 1.5~T and 1.0~T
axial magnetic fields, respectively.

In both studies we demand the presence of only the transition pion
(via $\pizgg$) in hadronic events, and impose no restriction on the $\hbc$
decay other than the global event selections described later.  Branching
fractions for $\hbc\to\gamma\etabc$ are known to be large, so that $\hbc$
searches can and do reasonably seek to require either the radiative
photon and/or a reconstructed $\etabc$; we avoid such an approach here.
Instead, the magnitude of the $\hbc$ signal is inferred from the size of any
enhancement in the distribution of 
$\recmas$, the mass recoiling against the putative transition $\piz$
($\recmas\equiv \sqrt{(p_{res}-p_{\pi^0})
^2}$ where $p_{res}$ and $p_{\pi^0}$ are the initial $\Upsilon(3S)[\psi(2S)]$ 
and the $\pi^0$ four-momenta, respectively).
The main challenge is to design selection
criteria to simultaneously preserve signal strength
while suppressing the unavoidably large backgrounds.

We determine our selection criteria and fit procedure based 
on Monte Carlo (MC) simulations of resonance decays, 
continuum ($e^+e^-\to\gamma^*\to q\bar{q}$) background,
and an off-resonance sample, 20.7~pb$^{-1}$ taken 
$\sim$16\mevcc\ below the $\pp$, so as to avoid bias.  
The MC generation utilizes {\sc EvtGen}~\cite{evtgen},
the values shown in Table~\ref{tab:mc}, 
and the most recent relevant branching fraction, mass, and 
width measurements~\cite{Nups,chibcasc,PDG} and predictions
(where measurements do not exist), followed by a {\sc GEANT}-based
simulation~\cite{geant} of the two detector configurations.
First, global event selections for $\ut$ and $\pp$ are performed
(with 92\% and 95\% efficiencies)
as in Refs.~\cite{Nups} and \cite{psi2sincgam}, respectively.
In order to isolate reliably-triggered
resonance decays and suppress $e^+e^-$, $\mu^+\mu^-$, and $\tau^+\tau^-$
final states, the selections demand that there be well-reconstructed
charged particle tracks and that the total measured energy in each
event be consistent with that of the initial $e^+e^-$ state.
The copious background from $\psi(2S) \to \pi^+ \pi^- J/\psi$ 
is reduced by requiring that events have no oppositely-charged
pair of particles with dipion recoil mass near that of $J/\psi$.

\begin{table}
\caption{Four key values used in each MC-based
optimization of selection criteria
and some related energies of interest.
\label{tab:mc}}
\setlength{\tabcolsep}{0.82pc}
\begin{center}
\scalebox{1.2}
{
\begin{tabular}{lccc} \hline \hline
\rule[10pt]{-1mm}{0mm}
Item & Units &  $\ut$   &  $\pp$  \\[0.05mm]
\hline
\rule[10pt]{-1mm}{0mm}
$M(\hbc)$           & \mevcc     & 9900.0 & 3525.28 \\[0.40mm]
$\Gamma(\hbc)$      & \mevcc     & 0      & 0.86    \\[0.40mm]
$\bhbcgebc$         & \%         & 38.0   & 37.7    \\[0.40mm]
$\brf(\piz\hbc)$    & $10^{-4}$  & 16     & 8.4     \\[1.60mm]
$E_\piz(\piz\hbc)$  & MeV        & 446    & 159     \\[0.40mm]
$E_\gamma(\gamma\chi_{b[c]2,1})$ & MeV    & 433, 452 & 128, 171\\[0.40mm] 
\hline \hline
\end{tabular}
}
\end{center}
\end{table}

We reconstruct the transition $\pizgg$ candidates based on pairs of 
showers in the calorimeter that are not matched to the projected 
trajectory of any charged particle. Showers must be located well 
within the boundaries of the crystal calorimeter barrel 
($|\cos \theta|<0.81$) for $\ppphc$, or barrel and endcaps ($0.85<|\cos \theta|
<0.93$) for $\utphb$, where $\theta$ is the polar angle with respect to the
positron beam direction.  
Also, the showers
must have $E_\gamma>30 (50)$\mev\ in the barrel (endcap)
calorimeter.
To reduce background from non-photon hadron-induced showers,
photon candidates are also required to have a lateral shower profile 
consistent with that of an isolated electromagnetic shower. 
Defining the $\pi^0$ mass ``pull'' 
($\equiv (M_{\tg}-M_{\pi^0})/{\Delta M_{\tg}}$,
where $\Delta M_\tg$ is the photon-energy-dependent
resolution on $\tg$ invariant mass, typically $\sim$5-7\mevcc),
we find that restricting the mass pull to [$-3.0$, $+2.5$] optimizes 
sensitivity to signal. The asymmetric mass pull range accounts for 
the presence of a low-side tail in $M_\tg$ caused by lateral and 
longitudinal shower leakage from the crystals assigned to the photon 
candidate's shower. If a daughter photon is shared with more than one 
$\pi^0$ candidate, the pair with smaller mass pull is chosen.
We then kinematically constrain $M_{\tg}$ to the known $M_{\pi^0}$ to
improve $\pi^0$ momentum resolution.

By the very nature of such an inclusive measurement, most of our selected
events will be background, and any inference of an $\hbc$ signal depends 
strongly on two characteristics of the background: first, that it has 
smooth $\recmas$ dependence in the vicinity of the $\hbc$ mass, 
so that extrapolation of its shape underneath any signal can be made with 
confidence, and second, that its magnitude can be reduced enough 
to observe a peak of adequate statistical significance. Enormous 
combinatoric smooth backgrounds are present from $\pizgg$ and $\etagg$,
which can arise from $e^+e^-\to q\bar q$ or at various
stages of resonance transitions or decays. 
Although without structure in $\recmas$, these must be suppressed 
without sacrificing too much signal. Moreover, significant non-smooth 
structures in $\recmas$ are necessarily present as well. Because the 
spin-singlet masses are 
near the spin-weighted averages of the respective spin-triplet states, 
the transition-$\piz$ energy is always close to the photon energies from 
the electric-dipole (E1) transitions $\utppgchibcj$,
as shown in Table~\ref{tab:mc}.
An E1 photon will frequently be paired with one of the multitude of 
very low energy photons in most events to form a mass close to that 
of a true $\piz$. As the above-mentioned E1 photons are monochromatic, 
non-smooth contributions in $\recmas$ can arise near the $\hbc$ 
mass. 
These sources of fake $\piz$ candidates make extrapolating a reliably 
known background shape underneath 
any $\hbc$ signal systematically challenging unless mitigating measures 
are taken. 

\begin{figure}[t]
\begin{center}
\includegraphics[width=1.0\textwidth]{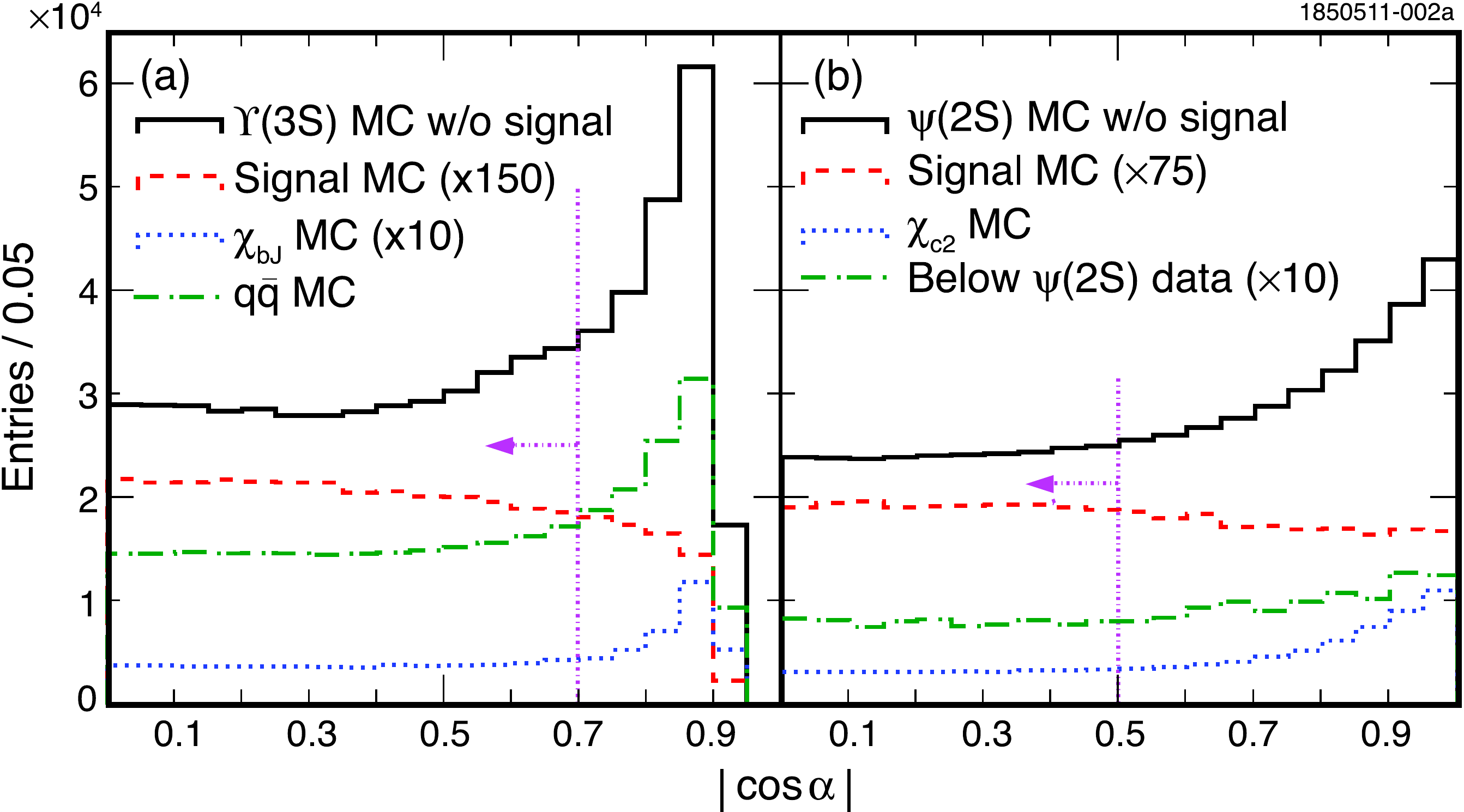}
\end{center}
\caption{
Distributions of the $\piz$ decay angle $\alpha$
for MC (or data, where noted below) samples scaled to the size of our datasets:
(a) $\utphb$ signal (dashed), all $\ut$ MC decays except signal (solid), 
$\utgchibj$ (dotted), and $e^+e^-\to\gamma^*\to q\bar q$ (dash-dot),
for $9875<\recmas<9925$\mevcc;
(b) $\ppphc$ signal (dashed), all $\pp$ decays except signal (solid), 
$\ppgchicj$ (dotted), and below-$\pp$ continuum data (dash-dot),
for $3520<\recmas<3530$\mevcc.
Vertical lines and arrows show selected regions of $\aca$.
}
\label{fig:fig1}
\end{figure}

\begin{figure}[t]
\begin{center}
\includegraphics[width=1.0\textwidth]{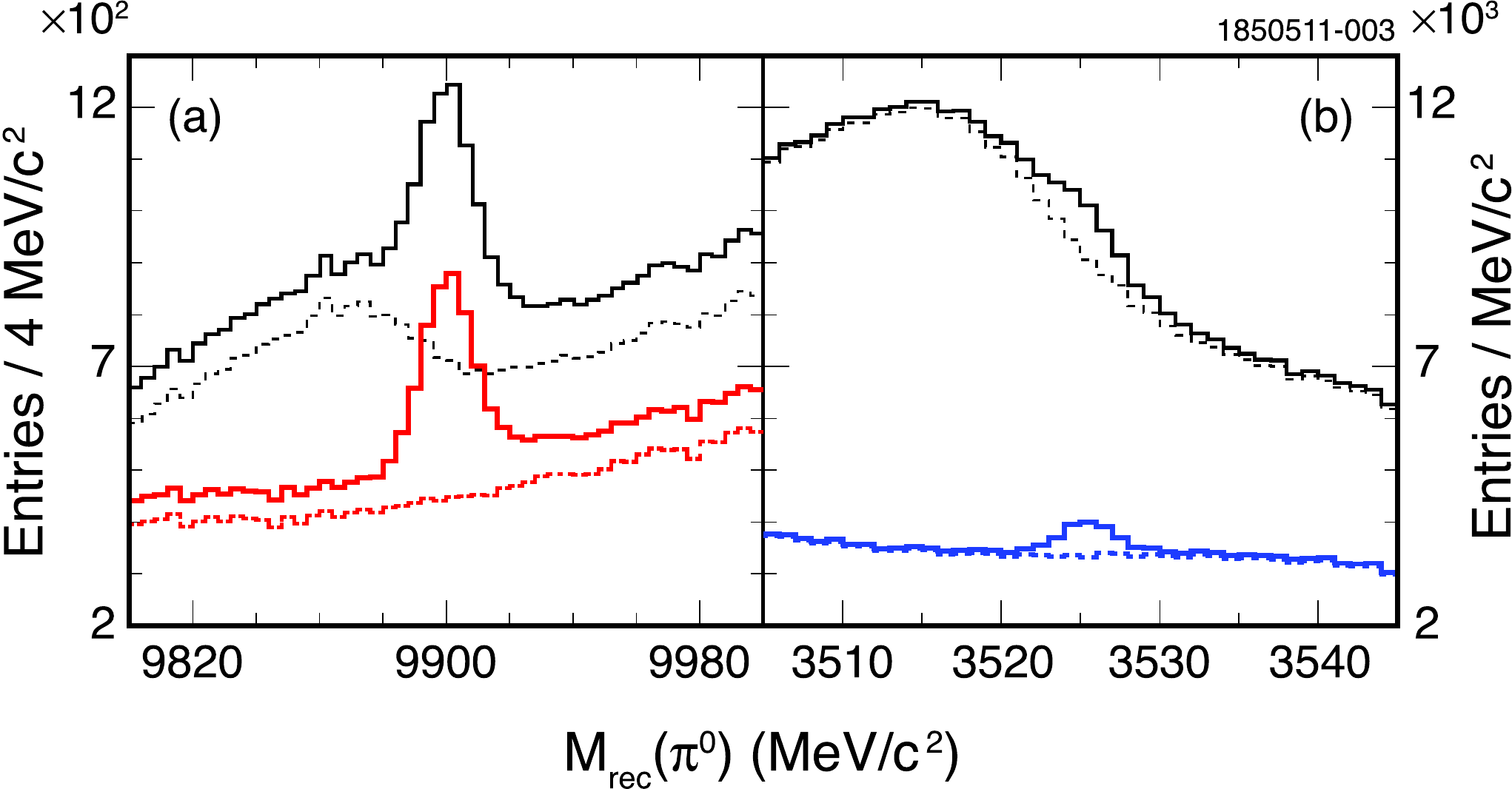}
\end{center}
\caption{
Solid histograms show
distributions in $\recmas$ for MC samples of 
signal $\piz\hbc$ transitions at modeled levels 
plus (a) $\gamma\chibj$, and (b) $\gamma\chict$ backgrounds,
scaled to the size of our datasets, 
with our 
restrictions on $\piz$ 
decay angle 
[(a) $\aca<0.7$, (b) $<0.5$] (lower pairs of 
histograms)
and with $\aca<1.0$ (upper pairs). Dashed
histograms show the contribution of (a) $\gchibj$,
and (b) $\gchict$-only events
for both selections on $\aca$.
\label{fig:fig2}}
\end{figure}

Our chosen method for fake-$\piz$ background suppression is to restrict 
the values of the $\pizgg$ decay angle $\alpha$, taken as the 
angle in the $\piz$ center-of-mass frame between either photon and 
the $\piz$-boost direction.  True $\piz$ decays have a uniform distribution in
$\aca$.  Values of $\aca$ near unity imply an asymmetric decay, with one of the
photons being very soft; it is these candidates which give rise to most of the
background.  MC studies indicate that $\aca<0.7~[0.5]$ provides the best
compromise between sensitivity and background rejection for $\Upsilon(3S) \to
\pi^0 h_b$ [$\psi(2S) \to \pi^0 h_c$].  The tighter value for
$\ppphc$ reflects the order-of-magnitude larger values of
$\bppgchicj/\bppphc$ relative to $\butgchibj/\butphb$ and
the consequent need to suppress the $\gamma\chicj$ backgrounds
more severely. All these points are demonstrated in Figs.~\ref{fig:fig1}
and \ref{fig:fig2}, where it can be seen that backgrounds
congregate at larger $\aca$ and cause structure in $\recmas$ unless
suppressed.
The signal histograms in Fig.~\ref{fig:fig1} fall with increasing $\aca$
due to the requirement that $E_\gamma>30 (50)$\mev\
 in the barrel (endcap) calorimeter, the $\recmas$ range
 restriction for these plots, and other event selection criteria.
Exclusion of asymmetric $\pizgg$ decays also has the advantage of
improving $\recmas$ resolution (which enhances sensitivity) because 
such decays include softer photons, which have poor relative energy resolution.

\begin{table}
\caption{Features of the fits for signal extraction.
``Order'' in background shapes refers to order of polynomials.
See text for other details.
\label{tab:fits}}
\setlength{\tabcolsep}{0.33pc}
\begin{center}
\scalebox{1.2}
{
\begin{tabular}{lcc} \hline \hline
\rule[10pt]{-1mm}{0mm}
Item &  $\ut$   &  $\pp$  \\[0.05mm]
\hline
\rule[10pt]{-1mm}{0mm}
$\recmas$ binning   & 4\mevcc            & 1\mevcc                 \\[0.40mm]
$\recmas$ fit range & 9.8-10.0\gevcc     & 3505-3545\mevcc         \\[0.40mm] 
Background shape    & $3^{\rm rd}$-order & ARGUS+$2^{\rm nd}$-order\\[0.40mm]
$M(\hbc)$ for fit   & 9900\mevcc         & 3525.42\mevcc           \\[0.40mm]
Signal shape        & Reversed CBL       & Double Gaussian         \\[0.40mm]
\hline \hline
\end{tabular}
}
\end{center}
\end{table}

Signal extraction is accomplished by fitting each $\recmas$ 
distribution with suitable binning and range for the
combination of smooth background and peaking signal 
component shapes with fixed $M(\hbc)$ and floating normalizations for each; 
choices of mass, binning, range, and shapes appear in Table~\ref{tab:fits}. 
The $\hb$ signal shape is found to be best represented by a reversed Crystal
Ball line (CBL) shape (an ordinary CBL shape~\cite{CBL} with the power-law tail 
on the high side) while fixing the shape parameters based on signal-only fits
to MC samples.  A third-order polynomial for the $\hb$ background fits the data
well.  The $\hc$ shape is chosen instead as a double Gaussian with independent
means (the mean of the broader Gaussian is shifted higher)
based on studies of our signal MC sample.
The different signal shapes reflect those expected for very 
slow ($\hc$) and faster ($\hb$) $\piz$ mesons, and reflect that the
calorimeter resolution is more symmetric at lower energies
and develops a low-side tail at higher energies due to shower leakage;
the tail moves to the high side in recoil mass.
Since $E_\piz$ from $\psi(2S)\to\piz\hc$ 
is close to the kinematic limit of $M(\piz)$, 
the $\hc$ background shape is represented by an ARGUS 
function~\cite{ARGUS}, which effectively models reduced phase space 
as a function of increasing mass, plus a second-order polynomial.
Our MC studies show we can extract branching fractions consistent
with what we input to the MC samples with these fit procedures.

\begin{figure}[t]
\begin{center}
\includegraphics[width=1.0\textwidth]{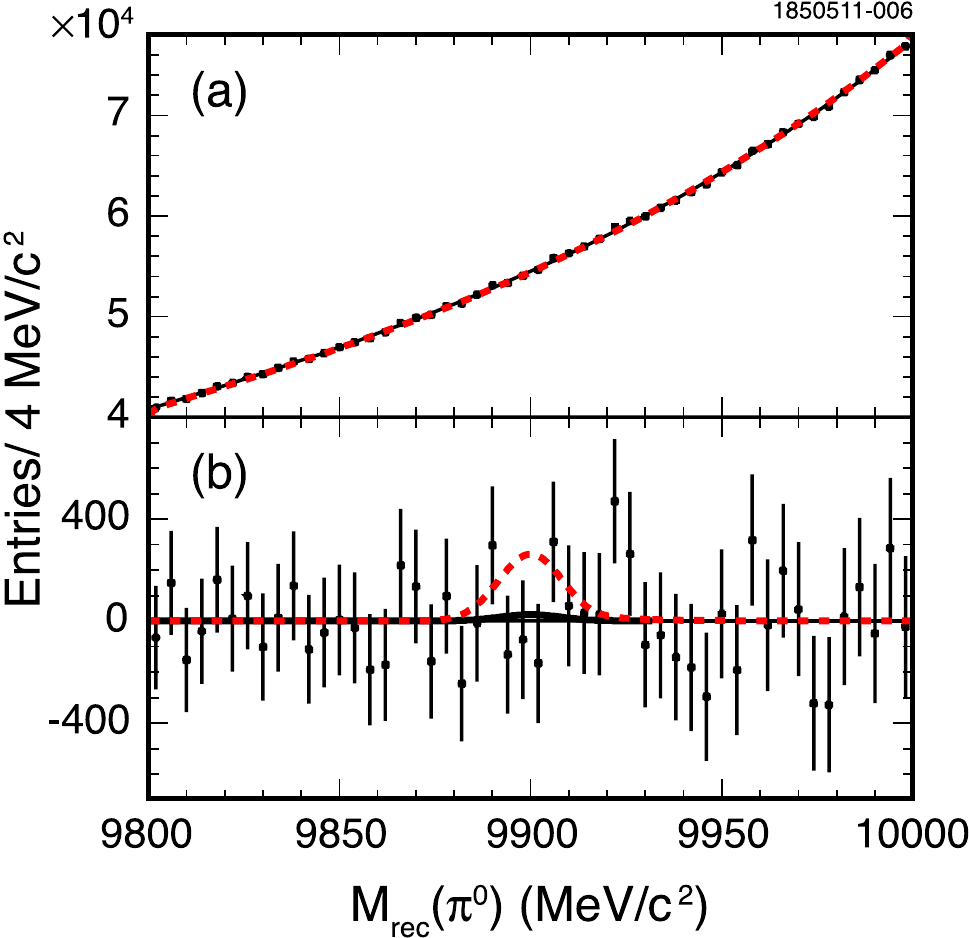}
\caption{
(a) Fit to $\recmas$ for $\utphb$ for fixed $M(\hb)=9900.0$\mevcc.
The $\chi^2$ value from this fit is 26.4 for 50 data points 
(minus 5 parameters) with confidence level of 98.8\%.
(b) The fitted background-subtracted spectrum (solid curve).
The dashed curve corresponds to the upper limit on signal candidates
at 90\%~CL  ($<1439$ events or
$\brf(\Upsilon(3S)\to\pi^0h_b)<11\times10^{-4}$ at 90\% CL).}
\label{fig:fig3}
\end{center}
\end{figure}

\begin{figure}[t]
\begin{center}
\includegraphics[width=1.0\textwidth]{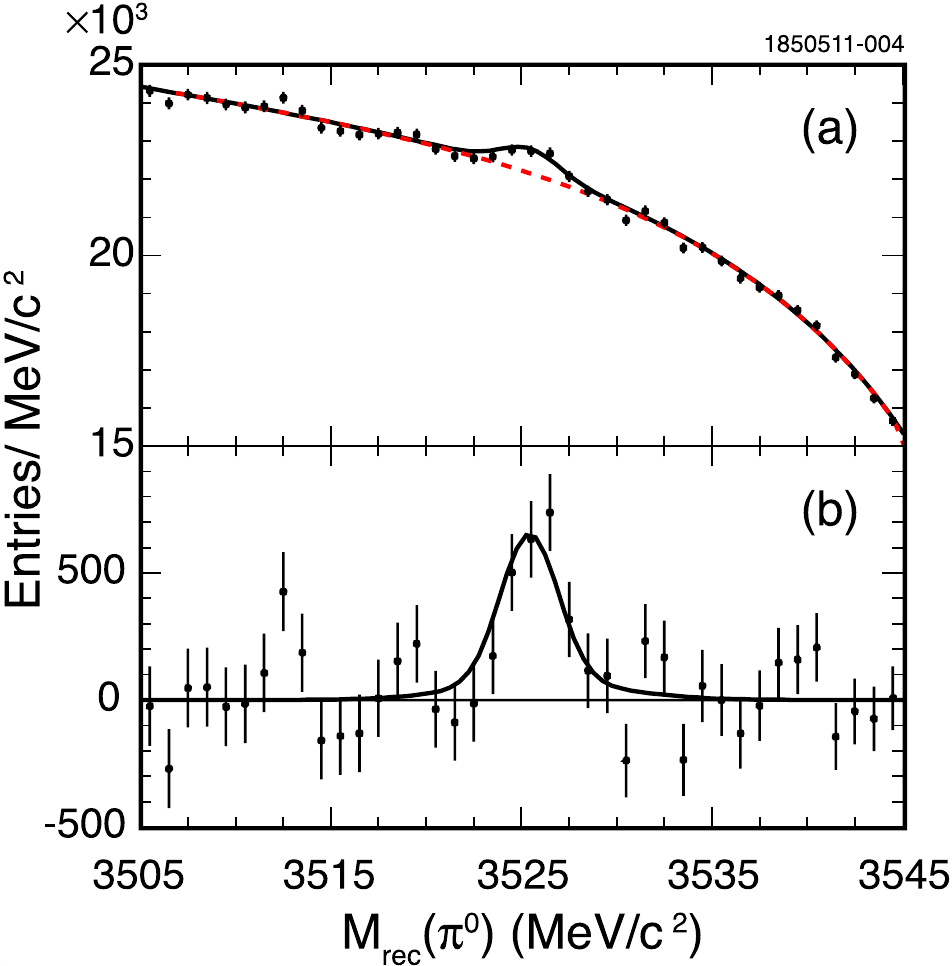}
\end{center}
\caption{(a) Fit to the $\ppphc$ $\recmas$ data 
with (solid curve) and without (dashed) a signal.
The $\chi^2$ value from this fit is 41.0 for 40 data points
(minus 5 parameters) with confidence level of 22.5\%.
(b) As in (a) but with the background fit from (a) subtracted.
\label{fig:fig4}}
\end{figure}

Results from the fits appear in Figs.~\ref{fig:fig3} and \ref{fig:fig4} and 
Table~\ref{tab:results}.  No signal is seen for $\utphb$, and the quoted upper
limit integrates over physical branching fractions only and includes systematic
effects (see below).  An unambiguous signal is observed for $\ppphc$.
The data points above the fit level near 3513\mevcc\ in Fig.~\ref{fig:fig4}(b)
have a width narrower than the detector resolution, and hence
must constitute a statistical fluctuation.
The statistical significances shown are computed as
$\sqrt{-2\ln{(L_{\rm wo}/L_{\rm w})}}$, where 
$L_{\rm wo}$ and $L_{\rm w}$ are likelihood values from
fits of $\recmas$ without and with signal shape components, 
respectively. If the $\hc$ mass is allowed to float,
$3525.9\pm0.3$\mevcc
(a mass value $1.1\sigma$ larger than the world average~\cite{PDG}) is obtained
which results in negligible change in fitted yields with respect to
the case when the mass is fixed to the world average.
The final branching fractions are obtained as
$\brf = \Nevt/(\epsilon\Nres)$ where $\Nevt$ is the 
number of signal events extracted from the fit, $\Nres$ 
is the number~\cite{Nups,npsi2s} of resonance decays in the dataset, and 
$\epsilon$ is the reconstruction efficiency obtained from 
performing similar fits on MC samples.

We consider a variety of sources for systematic
uncertainties on the branching fractions obtained
from the fits, and summarize them in
Table~\ref{tab:sys} along with estimates of
their contributions. The general approach
is to vary the important selection criteria
or fitting choices over reasonable ranges 
and note any resulting variations in
$\butppphbc$ beyond expected statistical changes, 
and then to add all such
effects in quadrature. The dominant systematic
effects are quite different for $\hb$ and $\hc$.
For $\hb$, the 
source that stands out is
the $\recmas$ 
fit range.
For $\hc$, the largest contributions
come from
the functional form of the background shape,
uncertainty in $\Gamma(\hc)$,
and our understanding of $\piz$ resolution
in data and MC simulation.
As $\recmas$ resolutions are larger
than the $\hbc$ mass uncertainties~\cite{BaBarhb,Bellehb,BEShc,PDG},
there is no need for separate systematic 
errors from such variation.

\begin{table}
\caption{Signal-extraction efficiencies and final measured 
event yields and branching fractions, the 
latter including systematic uncertainties. 
The upper limit integrates over physical values only.
See text for details.
\label{tab:results}}
\setlength{\tabcolsep}{0.60pc}
\begin{center}
\scalebox{1.2}
{
\begin{tabular}{lcc} \hline \hline
\rule[10pt]{-1mm}{0mm}
Item &  $\ut$   &  $\pp$  \\[0.05mm]
\hline
\rule[10pt]{-1mm}{0mm}
$\epsilon(\piz\hbc)$       & 21.3\%           & 12.6\%           \\[0.40mm]
$\Nevt$                    & $139\pm821$     & $2943\pm501$     \\[0.40mm]
Significance               & $0.2\sigma$      & $5.9\sigma$      \\[0.40mm]
$\Nres(10^6)$             & $5.88\pm0.12$    & $25.9 \pm 0.5$ \\[0.40mm]
$\brf(\piz\hbc)~(10^{-4})$ & $<12$~at 90\%~CL   & $9.0\pm1.5\pm1.3$\\[0.40mm] 
\hline \hline
\end{tabular}
}
\end{center}
\end{table}
For the $\hb$ $\recmas$ fit range, we consider alternate ranges 50\mevcc\ 
wider and narrower, symmetrically around our 
chosen $M(\hb)$, observing excursions
as large as noted in Table~\ref{tab:sys}.
For background shape, a fourth-order polynomial
is tried instead of a third-order; for signal
shape, we allow a double Gaussian instead 
of the reversed CBL shape (effects of 
imperfectly understood $\recmas$ resolution are 
also accounted for in this variation).
To test the dependency of our result upon the predicted 
resolution in $\recmas$, we decrease the smearing predicted
by the MC by 8.5\% less than predicted (which no longer
gives reasonable agreement between data and MC samples),
and consider half of the change in its measured branching fraction
as a possible systematical bias.
For binning of $\recmas$, we vary from 4 to both 2 and 6\mevcc\ for $\hb$.
We allow $\bhbgeb$ to vary from 0\% and up to 100\%
because its size is unknown and has a small but 
nonzero effect on photon multiplicity and therefore
upon the efficiency of the signal $\pi^0$ reconstruction.
We also account for the uncertainty in $\Nres$.

For the $\ppphc$ fit, the background shape is alternately fit to either first-
or third-order polynomials instead of the second-order.  The $\hc$ signal size
appears to have an approximately linear dependence on the assumed
$\Gamma(h_c)$, behaving as 
$\bppphc = [7.6 + 1.4\Gamma(\hc)/\Gamma_0] \times 10^{-4}$,
where $\Gamma_0=0.86$\mevcc\ is the chosen width.  We then vary the width by
$\pm50$\% of $0.86$\mevcc\ to estimate a systematic error.
We account for uncertainty in calorimeter resolution by varying it over ranges
that still represent the data reasonably well, as in our $h_b$ study.
We vary $\acamax$ ($0.5\pm0.1$), the fit range (3505--3550\mevcc; 
we could not go lower than
3505\mevcc\ due to the contamination from $\ppgchict$), 
suppression of events from $\pp\to\pi^+\pi^-J/\psi$, 
bin width ($1.0\pm0.5$\mevcc), as well
as doubling and halving the assumed $\bhcgec$, which includes
the value measured by BESIII~\cite{BEShc}.

\begin{table}
\caption{Summary of relative systematic uncertainties in percent on 
$\butphb$ and $\bppphc$.  Entries marked ``$\cdot\cdot\cdot$'' make negligible contributions.
\label{tab:sys}}
\setlength{\tabcolsep}{1.3pc}
\begin{center}
\scalebox{1.2}
{
\begin{tabular}{lrr} \hline \hline
\rule[10pt]{-1mm}{0mm}
Source &    $h_b$\ \  &  $h_c$\ \  \\[0.15mm]
\hline
\rule[10pt]{-1mm}{0mm}
Background shape           & 0.8  &  9.3  \\[0.35mm]
$\Gamma(\hbc)$             & $\cdot\cdot\cdot$    &  7.8  \\[0.35mm]
Fit range                  & 19.7 & $\cdot\cdot\cdot$     \\[0.35mm]
Binning                    & 10.9 & $\cdot\cdot\cdot$    \\[0.35mm]
Signal shape               & 1.7  & $\cdot\cdot\cdot$     \\[0.35mm]
$\pi^0$ resolution         & 2.0  &  6.6  \\[0.35mm]
$\Nres$                    & 2.0  &  2.0  \\[0.35mm]
$\bhbcgebc$                & 2.5  &  4.1  \\[0.35mm]
Efficiency (MC statistics) & 0.4  &  0.6  \\[0.35mm]
$\aca$                     & $\cdot\cdot\cdot$    & $\cdot\cdot\cdot$     \\[1.2mm]
Quadrature sum             & 22.9 &  14.6 \\
\hline \hline
\end{tabular}
}
\end{center}
\end{table}

 In conclusion, we have measured branching fractions
for $\utppphbc$ as shown in Table~\ref{tab:results}.
The $\hb$ upper limit is dominated by statistical
uncertainties and supersedes the previous CLEO limit,
$\brf[\Upsilon(3S)\to\pi^0h_b] < 2.7 \times 10^{-3}$ at $90\%$ CL
\cite{OldCLEO}.
If we combine the product branching fraction from
\babar~\cite{BaBarhb} with our result, considering only
physical values of branching fractions, we
infer $\bhbgeb>24$\% at 90\%~CL, consistent with predictions~\cite{hbmodel}.
The $\hc$ result is consistent 
with the value from BESIII~\cite{BEShc}.

\begin{acknowledgments}
We gratefully acknowledge the effort of the CESR staff in providing us with
excellent luminosity and running conditions. 
D. C.-H. thanks the
A.P.~Sloan Foundation.  This work was supported by the National Science
Foundation, the U.S. Department of Energy, the Natural Sciences and Engineering
Research Council of Canada, and the U.K. Science and Technology Facilities
Council.
\end{acknowledgments}


\begin{thebibliography}{99}

\bibitem{hqppo} N.~Brambilla {\it et al.},  Eur. Phys. J. C {\bf 71}, 1534 (2011).

\bibitem{CLEOhc} J. Rosner {\it et al.} (CLEO Collaboration), Phys.\ Rev.\
Lett.\ {\bf 95}, 102003 (2005); P. Rubin {\it et al.} (CLEO Collaboration),
Phys.\ Rev.\ D {\bf 72}, 092004 (2005); S. Dobbs {\it et al.} (CLEO
Collaboration), Phys.\ Rev.\ Lett.\ {\bf 101}, 182003 (2008).

\bibitem{BEShc} M. Ablikim {\it et al.} (BESIII Collaboration),
Phys.\ Rev.\ Lett.\ {\bf 104}, 132002 (2010).

\bibitem{BaBarhb} J. P. Lees {\it et al.} (\babar~Collaboration),
arXiv:1102.4565v2.

\bibitem{pipihc}
T.K. Pedlar {\it et al.} (CLEO Collaboration),
 Phys.\ Rev.\ Lett.\ {\bf 107}, 041803 (2011).

\bibitem{Bellehb} I. Adachi {\it et al.} (Belle Collaboration),
arXiv:1103.3419.

\bibitem{hbprod}  M. B. Voloshin, Yad.\ Fiz.\ {\bf 43}, 1571 (1986) 
[Sov.\ J. Nucl.\ Phys.\ {\bf 43}, 1011 (1986)];
M.~B. Voloshin and V.~I. Zakharov, Phys.\ Rev.\ Lett.\ {\bf 45}, 688 (1980).

\bibitem{hbprivate} U.G. Mei{\ss}ner, private communication.

\bibitem{hbmodel} S.~Godfrey and J.~L. Rosner,
Phys.\ Rev.\ D {\bf 66}, 014012 (2002).

\bibitem{Kuang:2002hz}
  Y.~P.~Kuang,
 Phys.\ Rev.\  D {\bf 65}, 094024 (2002).

\bibitem{Guo:2010zk}
  F.~-K.~Guo, C.~Hanhart, G.~Li, U.~-G.~Meissner, Q.~Zhao,
 Phys.\ Rev.\ D  {\bf 82}, 034025 (2010).

\bibitem{Burns:2011fu}
  T. J. Burns, arXiv:1105.2533 [hep-ph].

\bibitem{Nups} M. Artuso {\it et al.} (CLEO Collaboration),
Phys.\ Rev.\ Lett.\ {\bf 94}, 032001 (2005).

\bibitem{npsi2s} H. Mendez {\it et al.} (CLEO Collaboration),
Phys.\ Rev.\ D.\ {\bf 78}, 011102 (2008).

\bibitem{CLEO2} Y. Kubota {\it et al.} (CLEO Collaboration), Nucl.\ Instr.\
Meth.\ Phys.\ Res.\, Sect. A {\bf 320}, 66 (1992).

\bibitem{CLEO3trk} D.~Peterson {\it et al.}, Nucl.\ Instrum.\ Meth.\ A
 {\bf 478}, 142 (2002).

\bibitem{CLEOc} R.A. Briere {\it et al.} (CLEO-c/CESR-c Taskforces \&
  CLEO-c Collaboration), Cornell LEPP preprint CLNS 01/1742 (2001)
  (unpublished),
http://www.lns.cornell.edu/public/CLNS/\\
2001/CLNS01-1742/cleocyb.pdf.

\bibitem{evtgen} D.J. Lange, Nucl. Instrum. Methods Phys. Res., Sect. A {\bf 462}, 152 (2001). 

\bibitem{PDG} K. Nakamura {\it et al.} (Particle Data Group), J. Phys.\ G
{\bf 37}, 075021 (2010).

\bibitem{chibcasc} M. Kornicer {\it et al.} (CLEO Collaboration),
Phys.\ Rev.\ D {\bf 83}, 054003 (2011).

\bibitem{geant} R. Brun {\it et al.}, 
  {\textsc Geant} 3.21, CERN Program Library Long Writeup W5013 (1993), unpublished. 

\bibitem{psi2sincgam} S. B. Athar {\it et al.} (CLEO Collaboration),
Phys.\ Rev.\ D.\ {\bf 70}, 112002 (2004). Here, however, 
the maximum deposited energy in the calorimeter, $E_{cal}$, is
loosened from $85\%$ of the center-of-mass energy ($E_{CM}$) to $100\%$.

\bibitem{CBL} J. E. Gaiser, Ph. D. Thesis, SLAC-R-255 (1982) (unpublished);
T. Skwarnicki, Ph. D. Thesis, DESY-F31-86-02 (1986) (unpublished).


\bibitem{ARGUS} H. Albrecht {\it et al.} (ARGUS Collaboration), Phys.\ Lett.\
B {\bf 241}, 278 (1990).

\bibitem{OldCLEO} F. Butler {\it et al.} (CLEO Collaboration), Phys.\ Rev.\
D {\bf 49}, 40 (1994).

\end{thebibliography}
\end{document}